\documentclass[conference]{IEEEtran}
\IEEEoverridecommandlockouts
\usepackage{cite}
\usepackage{amsmath,amssymb,amsfonts}
\usepackage{algorithmic}
\usepackage{acronym}
\usepackage{graphicx}
\usepackage{svg}
\usepackage{textcomp}
\usepackage{xcolor}
\def\BibTeX{{\rm B\kern-.05em{\sc i\kern-.025em b}\kern-.08em
    T\kern-.1667em\lower.7ex\hbox{E}\kern-.125emX}}
\usepackage[hidelinks]{hyperref}

\newacro{CDB}{CyEntEE Database}
\newacro{BEV}{Battery Electric Vehicle}
\newacro{BES}{Battery Electric Storage}
\newacro{EHP}{Electric Heat Pump}
\newacro{PV}{Photovoltaic Plant}
\newacro{SM}{Smart Meter}
\newacro{FMI}{Functional Mock-up Interface}
\newacro{FMU}{Functional Mock-up Unit}
\newacro{DHN}{District Heating Network}
\newacro{NHL}{Nominal Heat Load}
\newacro{SIES}{Smart Integrated Energy Systems}
\newacro{ICT}{Information and Communication Technology}
\newacro{ERM}{Entity Relationship Model}

\begin{document}

\title{Towards a more comprehensive open-source model for interdisciplinary smart integrated energy systems
\thanks{Funding: This research was funded by the I\textsuperscript{3}-Programme of the Hamburg University of Technology (TUHH) in the project `CyEntEE' and the Federal Ministry for Economic Affairs and Climate Action in the project `EffiziEntEE' under the project number 03EI1050A.}
}

\author{
    \IEEEauthorblockN{
        Béla Wiegel,
        Tom Steffen,
        Davood Babazadeh and
        Christian Becker
    }
    \IEEEauthorblockA{
        Hamburg University of Technology \\
        Institute of Electric Power and Energy Technology \\
        Hamburg, Germany \\
        Email: \{bela.wiegel, tom.steffen, davood.babazadeh, c.becker\}@tuhh.de
    }
}

\IEEEpubid{979-8-3503-3682-5/23/\$31.00~\copyright2023 IEEE}

\maketitle

\begin{abstract}
The energy transition has recently experienced a further acceleration. In order to make the integration of renewable energies as cost-effective, secure and sustainable as possible and to develop new paradigms for the energy system, many energy system models have been developed in research in the past to evaluate the solutions. While model identification and dissemination of results are widely discussed in the literature, a detailed view of the methodology is often missing. This paper addresses this topic and proposes a methodology to build a comprehensive, publicly accessible database for modeling a multi-modal integrated energy system. The focus hereby is dynamic modeling of low- and medium-voltage grids consisting of prosumers, battery storages, heat pumps and electric cars. In addition, a district heating network is parameterized to match the electricity grid. Modelica and the TransiEnt-Library serves as the modeling tool. The methodology for creating the grid models is available via GitLab. A study case that uses the methodology to analyze the congestion situation within a medium-voltage distribution grid is presented.
\end{abstract}


\section{Introduction}

    The transition towards a modern and carbon-free energy system accelerates worldwide. While the shutdown of single centralized conventional power plants and their substitution by wind and photovoltaic power plants is feasible in the current system, the comprehensive enrollment of renewable energies requires a paradigm change in construction and operation concepts of the future energy system~\cite{2020_Torbaghan_Flexibility_Dispatch_Cone_Relaxation}. Integrated energy systems, also known as multi-modal energy systems, couple the electricity, heat and gas grids together. This holds the possibility of multiple energy provisioning paths, reinforcing the system evolution. For the concept of \ac{SIES}, the complexity is additionally expanded by aspects of \ac{ICT} investigated as a component necessary for good functioning~\cite{2022_Babazadeh, 2021_Hoth_CyEntEE_Whitepaper}.
    
    In order to understand and support the holistic design of complex systems of this kind -- both in terms of structural and operational planning -- modeling is an important tool. Therefore, many different energy system models were developed in the past, representing lots of concepts and use cases. For each model to be developed, the research question must be posed in advance to clearly design the requirements and create the model-specific methodology.

    \IEEEpubidadjcol
    
    While research questions and gap identification is project-specific, the work also consists of recurring tasks. Despite macroscopic differences, modeling often uses similar approaches and algorithms to solve the research question. This already starts with data collection and preparation, and continues with model building and evaluation methods. So that recurring tasks in the multitude of research projects do not have to be tackled in future projects again, the open research initiative has formed in the research community to make details of data collection and model-specific methodologies and implementation freely available. In Germany, a association called the National Research Data Infrastructure (nfdi)~\cite{2021_Hartl_NFDI} with the sub-group \emph{nfdi4energy} has founded with the aim of endorsing the research cycle and utilizing community services~\cite{2023_Web_nfdi4energy}. Also the European EERAdata and EnerMaps projects address this topic. In the ERIGrid projects, harmonizing scenario development and model validation is promoted~\cite{2020_Raussi_ERIGrid_Scenarios}. A main aspect in context of developing and publishing open source methodologies is the FAIR principle: findability, accessibility, interoperability and reusability. These principles are facilitated in current research~\cite{2023_Deng_OpenEnergy_Brasil, 2022_Suesser_F1000_OpenEnergy}, and especially challenges in data\footnote{The importance of data is also emphasized during the annual global \emph{Love Data Week}.} provision using open source tools in~\cite{2022_Berendes_Review_Open_Source_Frameworks_Modeling}.
    
\begin{figure}[tb]
        \centering
        \includegraphics[width=0.8\linewidth]{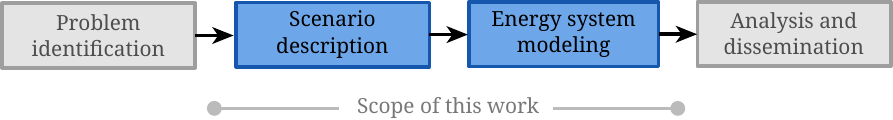}
        \caption{Scope of this paper within the research cycle.}
        \label{fig:intro_scope}
    \end{figure}

    In this context, the aim of this work is to develop a comprehensive energy system description in the form of a open source data and modeling framework that is suitable for the goal of generating detailed grid models with their node interconnecting energy conversion plants. In addition to modeling the electric power grid, a methodology is proposed to create a \ac{DHN} suitably parameterized to existing benchmark models for the electric power grid. While usually the dissemination of the modeling and research question investigation is the main part, this paper acts one step before, depicted in Fig.~\ref{fig:intro_scope}. Data and models are published and maintained via GitLab~\cite{2023_GitLab_CyEntEE_TUHH}. The main contributions are:
    \begin{itemize}
        \item Developing a comprehensive open-source database description for modeling of integrated energy systems at distribution level
        \item Creation of a \ac{DHN} appropriately parameterized to existing benchmark models for electricity grids
        \item Tool for generation of models for dynamic simulation of integrated distribution systems
    \end{itemize}
    Thus, the rest of the paper is organized as follows: section~\ref{sec:soa} gives an overview of existing approaches in public available modeling tools for integrated energy systems, section~\ref{sec:database_description} provides the data and modeling framework description and concepts for model creation which is applied in a use case in section~\ref{sec:use_case}, and section~\ref{sec:conclusion} gives final remarks and an outlook.
    
\section{State and challenges in open modeling of integrated energy systems} \label{sec:soa}

    Over the past decades, a large variety of energy system models have been created. Diverse requirements for the models in the complex environment of energy system research have led to manifold approaches. The ontology of models can be described with categories of spatial and temporal resolution, modeled energy carriers, physical detail, market representation and much more~\cite{2022_Fodstad_Modeling_Review_Challenges}. Over the years, constantly updated review articles classified the models and place them in context. Thereby, it becomes apparent that integrated energy systems, which couple different energy and application sectors, are becoming more and more important, especially in open source modeling~\cite{2022_Ouwerkerk_OpenSource_Models_Comparison}.
    
    In order to analyze the physical energy system, two main methods have been established in current research: optimization and simulation~\cite{2019_Groissboeck_Review_OpenSource_Models}. Optimization is suitable for planning a system configuration or an operational schedule that satisfies certain requirements. Objective functions with different weightings of the energy policy triangle consisting of economic efficiency, sustainability and reliability resp. resiliency are applied for this purpose. Methodologically, mixed-integer linear programming is often used here, or also dynamic programming~\cite{2019_Groissboeck_Review_OpenSource_Models}. Simulation aims to analyze time-dependent and mostly coupled processes and their interactions. In the case of quasi-stationary simulation, balance equations are used in which at each point in time the system is considered steady state. If storage-related time-dependent processes, e.g. in the form of an oscillation or transient processes within or across sectors are to be investigated, dynamic simulation in which transient processes are furthermore represented via numerical solution of differential equation systems are to be chosen. Only this approach gives insight in interactions between interconnected and controlled components. For optimization and simulation, it is necessary to capture the power system in the form of a mathematical model.
    
    In the planning and analysis of future energy systems, it is necessary in a preliminary step to design a system description in the form of energy system scenarios. Scenarios represent possible development paths and form the basis for the creation of the models used for simulation and optimization. Especially in interdisciplinary collaboration, which has many touch points in the planning phase, a common understanding of the energy system is of high importance so that consistent and concurring models can be created. In this context, multi-modality as a concept of the modern integrated energy system and its analysis requires the formation of grid benchmark models that are parameterized together with their energy conversion plants, such as heat pumps, combined heat and power or electrolyzers in order to be able to analyze their interactions during operation properly. In ERIGrid and ERIGrid 2.0, a EU-wide transnational program with lab-access and knowledge sharing in context of \ac{SIES}, concepts for harmonizing energy system validation approaches based on modeling and simulation are developed~\cite{2020_Raussi_ERIGrid_Scenarios}.

    Benchmark models for the electricity sector are established and well known. The SimBench project provides documented electricity grids and time series for load and consumption on all relevant voltage levels~\cite{2020_SimBench}, although some parts of the methodology are not fully explained. The Cigré benchmark system is another well established methodology and described in~\cite{2014_Cigre_Benchmark_Systems}. For the heat sector, different models for \ac{DHN} exist, e.g. in the \emph{pandapipes} library, for investigation of different research tasks, but there is no open benchmark model existing. This is accompanied by the lack of a benchmark system for the multi-modal resp. integrated energy system, which is a severe lack in literature. Taking this up, the following section describes a methodology to create a multi-modal test system for multi-modal grids, especially on distribution level.

\section{Open Database Description} \label{sec:database_description}

    The overall goal of this paper is to provide a freely available basis for modeling of lower-level integrated energy systems. Although the explained models are designed for dynamic simulation, it is also possible to use the same database to build linear optimization models, which was successfully tested. The special feature here is that the methodology as a whole is freely available, especially the data basis and the models itself, and large parts of the physical structure of the integrated energy system are covered. This is achieved by using the freely available TransiEnt-Library~\cite{2021_Senkel}, which contains component models of electricity, gas and heat grids and their most important coupling elements using differential-algebraic equation systems. The library is implemented in the modeling language Modelica~\cite{2021_Modelica}. With the dynamic energy system modeling, a bottom-up approach is chosen to gain a high degree of detail and represent the physical behavior of all components. The following described methodology takes this requirement as a basis.

    \subsection{Scenario Generation} \label{sec:ScenarioGeneration}
    
\begin{figure}[tb]
            \centering
            \includegraphics[width=0.8\linewidth]{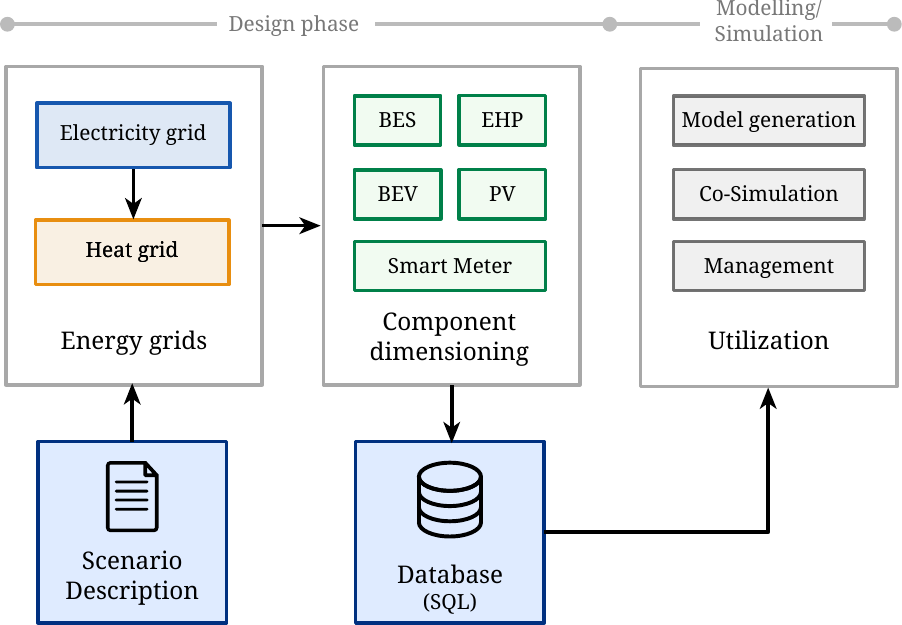}
            \caption{Methodology to create integrated test models. The scenario description is used to gain qualitative and -- where accessible -- quantitative information about the energy grid, which is quantified with all relevant components afterwards.}
            \label{fig:methodology_overall}
        \end{figure}

        The overall methodology for creating bottom-up energy system models on living quarter level is depicted in Fig.~\ref{fig:methodology_overall}. To obtain an overarching system description and prosumer parameterization, information on future energy systems is gathered in the form of a collection of different scenarios at first. Main sources for this are the TYNDP studies by european grid operators~\cite{2021_ENTSOG_ENTSOE_TYNDP2022_Final_Storyline_Report}, scientific studies~\cite{2021_DeAngelo_Nature_Energy_System_Scenarios}
        or results in form of exploratory scenarios of the before mentioned optimization models, e.g. REMod~\cite{2020_Sterchele_Fraunhofer_ISE_Scenarios}. The qualitative was former published in~\cite{2021_Hoth_CyEntEE_Whitepaper}. After this, representative grid topologies for the electricity grid are underlain since these are well established in research. The SimBench~\cite{2020_SimBench} benchmark model is the first chosen, but open-access Cigré or IEEE benchmark models are suitable as well. With a defined grid topology, the building structure is designed. Assumptions from the scenarios together with ISO standards -- or in case of Germany DIN standards -- are used as a basis for the parameterization of variables relevant for the model. Important variables here are the demand for electrical energy, the ground area of the building, thermodynamic properties such as heat transfer coefficients and their area fractions, and the resulting \ac{NHL} for the reference ambient temperature.
        
        With the defined building structure, a \ac{DHN} with its topology is parameterized. Since well established benchmark models for \acp{DHN} are scarce, a separate methodology was build. The following steps describe the procedure roughly: First, from different cycle-free grid topologies one is selected. Currently, a topology based on the power grid or a minimum spanning forest with the vertex spacing as weighting factor are available. Based on the \ac{NHL} values of each vertex and a comparatively low nominal supply temperature for modern \acp{DHN} in the nominal design case, the maximum required mass flow per vertex is calculated, which specifies the edge flows and thus the nominal width of the pipes at a defined maximum flow velocity.
    
\begin{figure}[tb]
            \centering
            \includegraphics[width=\linewidth]{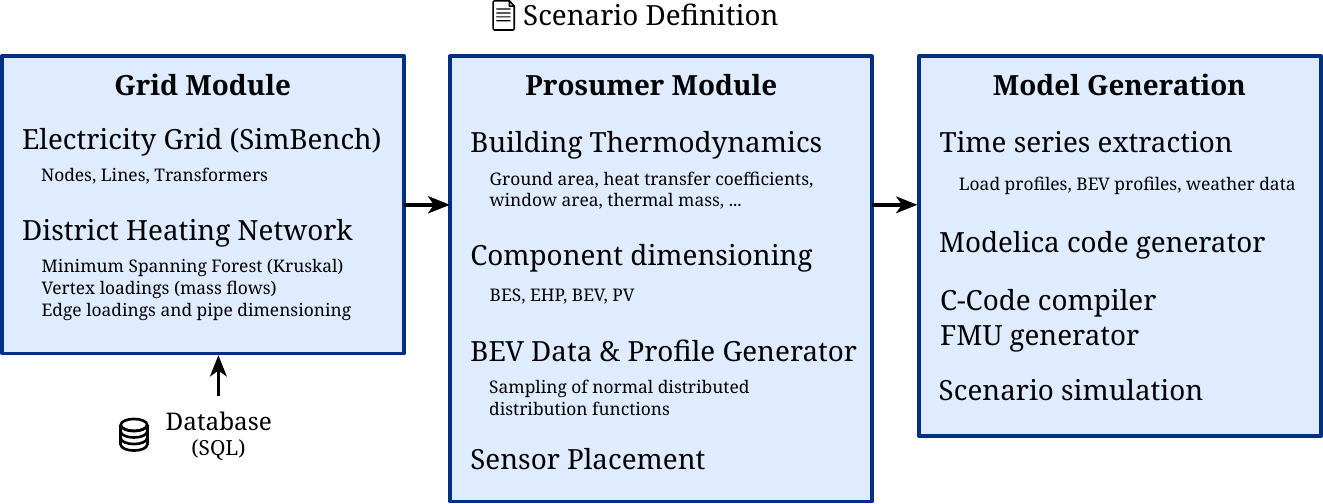}
            \caption{Scenario Generator: Overview of the modules contained in the repository for database and model creation.}
            \label{fig:methodology_detailed}
        \end{figure}

        All remaining components, especially \ac{BES}, \ac{EHP}, \ac{BEV}, \ac{PV} and properties for sensors like the \ac{SM} components are designed afterwards. Fig.~\ref{fig:methodology_detailed} gives an overview of the existing modules in the so called Scenario Generator, which constructs the scenario based on the qualitative scenario definition and different modules for the components of the modeled energy system. Technically, every aspects is implemented in single Python scripts, connected in a overall Scenario Generator script~\cite{2023_GitLab_CyEntEE_TUHH}. Scenario studies are possible this way by analyzing the effect of input data in the scenario description.

\subsection{Database description} \label{sec:database_description_technical}
    During the design phase mentioned before, all processed data is written into a database, called \ac{CDB}. In this case, a SQL database as back end is used, but any other back end like other database engines or file-based storages like JSON- or XML-files can be implemented. Advantage of a SQL database is the possibility for using in a co-simulation environment in which the modeling and simulation does not only take place on one single computer, but concurrently on multiple simulation platforms that need the same system information.
    
    The modeling database described was created as part of a project in a way that, in addition to modeling the physical energy system, also considers aspects of \ac{ICT} and decentralized markets~\cite{2021_Hoth_CyEntEE_Whitepaper} in form of a \ac{SIES}. In order to simulate the interactions between these different modeling entities, a co-simulation platform, which is part of current investigation, is developed.

    All components of the platform require data about the system configuration at simulation time. For this case, a relational database management system implementing SQL\footnote{In this case \emph{MariaDB} as open-source database management system.} was used as a back end to allow simultaneous read and write operations across multiple workstations, and always provide fast data access.

\begin{figure}[tb]
        \centering
        \includegraphics[width=0.6\linewidth]{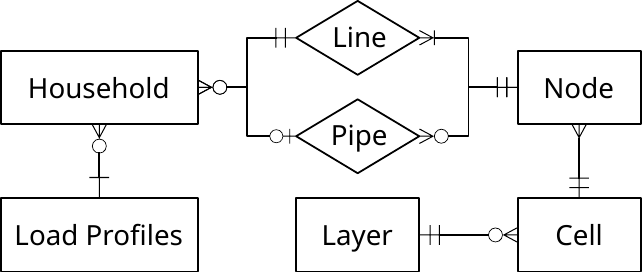}
        \caption{Entity Relationship Model for the table structure describing one living quarter. Each household, having a load profile, is connected by one line and by zero or one pipe to a node in the grid model, which itself is connected to a cell associated with a grid layer.}
        \label{fig:SQL_ERM}
    \end{figure}

\begin{figure}[tb]
        \centering
        \includegraphics[width=0.8\linewidth]{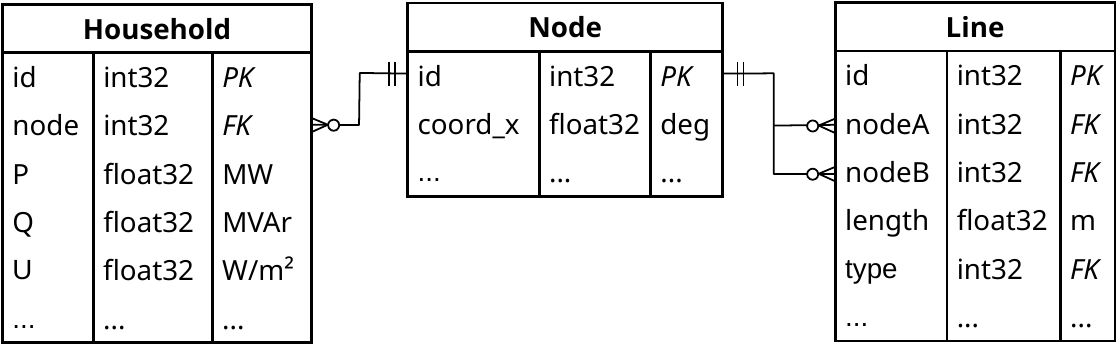}
        \caption{Exemplary entity relationship model for the relationship between the household, node and line table. Details about the table structure are given in the code~\cite{2023_GitLab_CyEntEE_TUHH}.}
        \label{fig:SQL_Entity_Example}
    \end{figure}

    In the case of the living quarter described in this context, the \ac{ERM} is shown in Fig.~\ref{fig:SQL_ERM}. The tabular structure is shown exemplary in Fig.~\ref{fig:SQL_Entity_Example}.

\subsection{Model Generation} \label{sec:database_model_generation}
    As stated before, the database holds the possibility to export and combine the scenario information in various different modeling environments, such as \textit{pandapower}, \textit{PyPSA}, \textit{Matlab} and various more. The simulation modeling approach with highest detail supported are dynamic models, which is choosen in this paper. Linear optimization models are also successfully implemented using the \ac{CDB}. This models, in contrast to quasi stationary or steady-state models, captures within time based simulations the time dependence of the state variables and their derivatives. The \ac{CDB} is prepared for the generation of those models. In order to make the conversion of the database contents into dynamic models particularly easy, a python module was implemented within the Scenario Generator, see Fig. \ref{fig:methodology_detailed}. This module effortlessly export and combine the scenario information within the \ac{CDB} into a model within the modeling language Modelica. Modelica is a language for modeling of cyber-physical systems, which supports acausal connection of components described by differential and algebraic equations \cite{2021_Modelica}.
    
    This module includes the logic to extract the realistic load profiles, study-based driving profiles, associated weather data as well as the topology including the detailed household specification from the database. The extracted information is then linked to existing and parameterizable Modelica models from the \textit{CyEntEE Models} package, which is planned to be released with this publication within the TransiEnt Library \cite{2021_Senkel}. 
    
    The model with the most significance in this package is the \textit{Prosumer} model, which is the representation of future households with the capability to at different points in time produce as well as consume power. The \textit{Prosumer} is described in higher detail in \cite{2022_Steffen}. This model consists in full configuration of six main components, a \ac{PV}, \ac{BES}, one or more \acp{BEV}, an \ac{EHP}, inflexible load and \ac{SM}. Additionally, the module offers to choose which of the elements \ac{BEV} and \ac{BES} on household level are externally controllable. In the model itself this is realized by means of \emph{RealInputs} from the Modelica Standard Library, which offer the external specification of target values while simulating. Furthermore, one can choose to enable the \ac{SM} models full potential meaning normal distributed Gaussian noise with no systematic deviation, known mean and known standard deviation, which is applied to the measurement outputs from the \ac{SM} model.

\begin{figure}[tb]
        \centering
        \includegraphics[width=90mm]{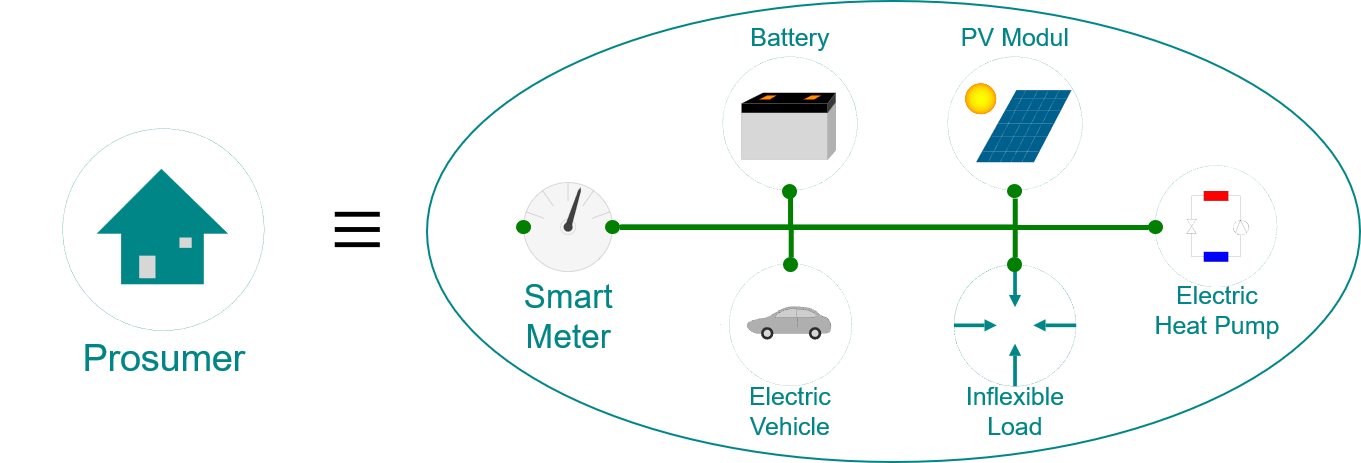}
        \caption{Structure of the Prosumer model from the \textit{CyEntEE Models} package.}
        \label{fig:Prosumer}
    \end{figure}

\begin{figure}[tb]
        \centering
        \includegraphics[width=90mm]{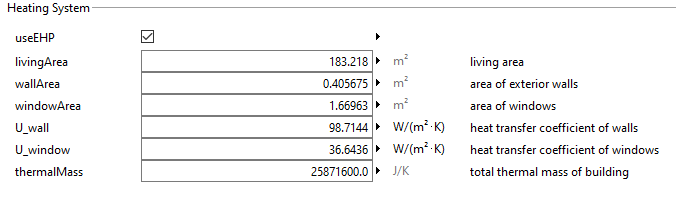}
        \caption{Example parameter configuration for Prosumer models household heating system from the \textit{CyEntEE Models} package.}
        \label{fig:Prosumer_Config}
    \end{figure}

\begin{figure}[tb]
        \centering
        \includegraphics[width=0.65\linewidth]{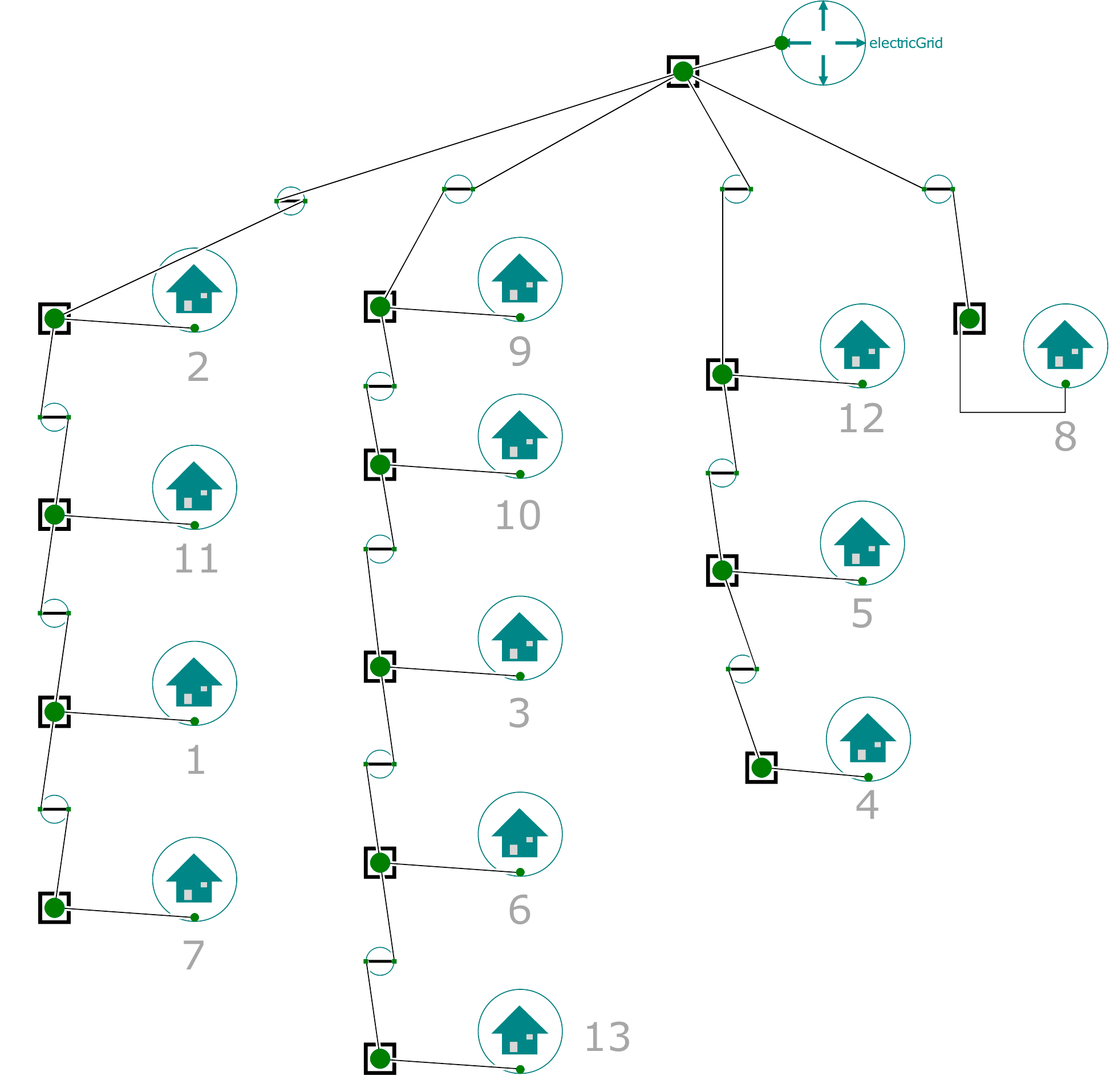}
        \caption{13-Household rural low voltage grid with topology based on SimBench \textit{LV-rural1--0-no\_sw}.}
        \label{fig:13HH-Grid}
    \end{figure}

    Fig. \ref{fig:Prosumer} and Fig. \ref{fig:Prosumer_Config} show the Prosumer models composition as well as the example parameter set for the configuration of household heating systems. The individually configured Prosumers are then integrated in grid structures, as mentioned in section \ref{sec:database_description}. The graphical representation of a simple 13 Prosumer low voltage grid in Dymola, which is a commercial modeling and simulation environment based on Modelica, is given in Fig. \ref{fig:13HH-Grid}. The grid model is based on the topology, meaning cable parameters, buses, household position and transformer, given by the SimBench dataset. The Prosumer configuration is obtained from the scenario generation, see section \ref{sec:ScenarioGeneration}, and is therefore, based on the stochastic realization of the model, different for each Prosumer. 
    
    \subsection{Model preparation for Co-Simulation} 
    According to the goal of the database to obtain common system descriptions usable for different kinds of research studies the preparation of dynamic models for Co-Simulation purposes was investigated and implemented. This is achieved by means of the \ac{FMI} Standard \cite{2021_Junghanns}. ``The Functional Mock-up Interface is a free standard that defines a container and an interface to exchange dynamic simulation models using a combination of XML files, binaries and C code, distributed as a ZIP file.'' \cite{2017_ModelicaAssociation}. The transformation from dynamic model within Modelica to the \ac{FMU} is again achieved by a python module within the Scenario Generator, see Fig. \ref{fig:methodology_detailed}.
    
\section{Use Case} \label{sec:use_case} 
\begin{figure}[tb]
    \centering
    \includegraphics[width=0.6\linewidth]{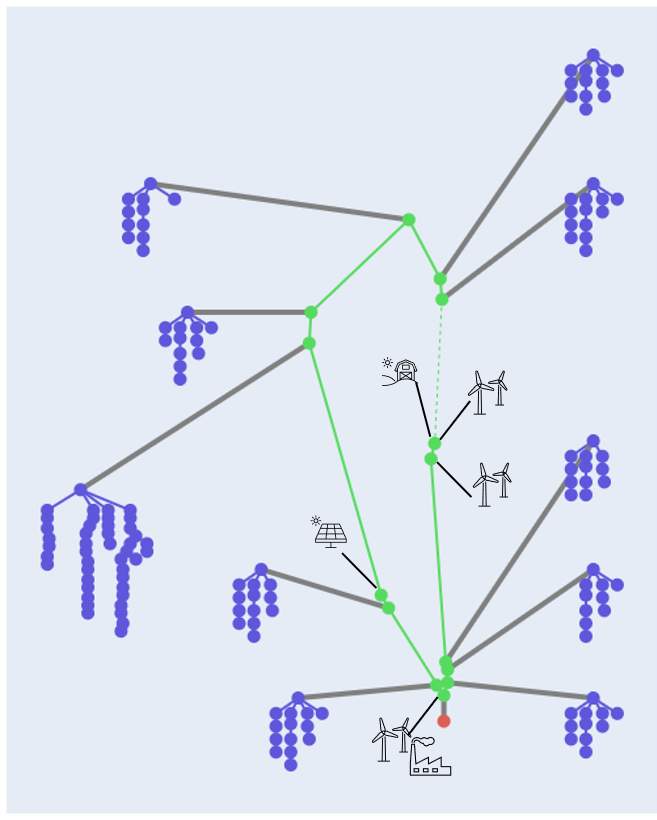}
    \caption{Rural medium voltage ring based on the SimBench \textit{MV-rural-2-no-switches} with detailed rural low voltage sub cells, generated from the SimBench \textit{LV-rural-1} and \textit{LV-rural-2}. Blue points are prosumers at low voltage level.}
    \label{fig:MV-Ring}
\end{figure}

To show the potential given by the database, a use case model in form of a rural medium voltage ring based on the SimBench \textit{MV-rural-2-no-switches} with detailed rural low voltage sub grids, generated from the SimBench \textit{LV-rural-1} and \textit{LV-rural-2} was generated within Modelica. Fig.~\ref{fig:MV-Ring} shows the schematic representation of this model, where the green grid represents the medium voltage ring with its lines and buses. The dark blue points and lines represent the buses within the underlying low voltage grids. They are connected by the grey lines, which represent a transformer with on spatial expansion. One bus, especially in the larger low voltage grid on the far left side, can connect more than one Prosumer to the grid. The single red point represents the interface, in form of an 630~MVA transformer to the high voltage grid which is in this simulation modeled as boundary condition.

The grids specifications, meaning share of \ac{BEV}, \ac{PV}  and other components, is based on the \emph{Distributed Energy Scenario} from \cite{2021_Hoth_CyEntEE_Whitepaper}. The weather data is taken from the Deutsche Wetterdienst (DWD) for Hamelin, a city in Lower Saxony in Germany, from the year 2020. The used load- and electric vehicle driving profiles are study-based generated and selected.

\begin{figure}[tb]
    \centering
    \includegraphics[width=1.0\linewidth]{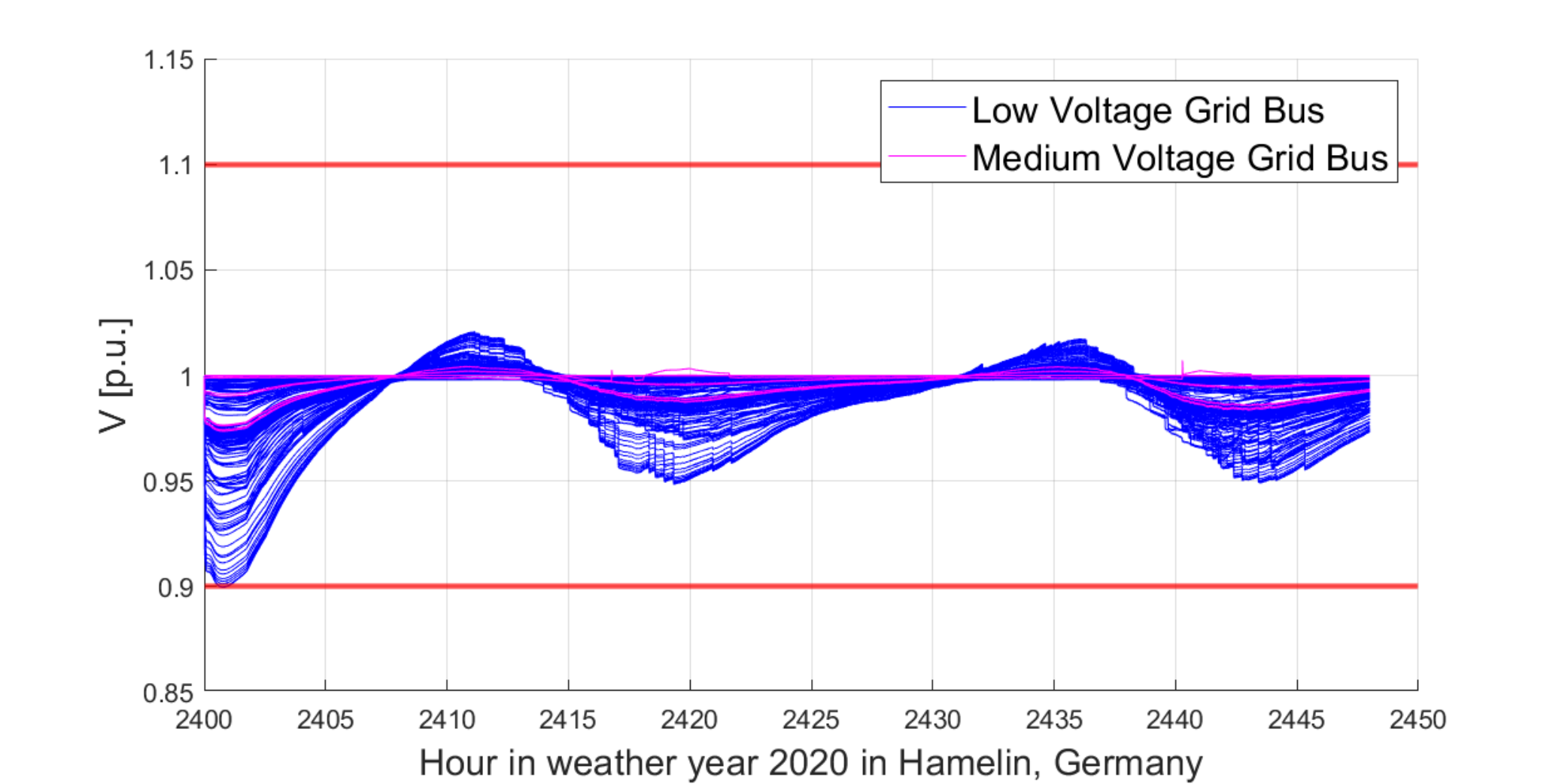}
    \caption{Voltages in per unit at low and medium voltage buses in the rural medium voltage ring scenario.}
    \label{fig:MV-Ring-Voltage}
\end{figure}

\begin{figure}[t]
    \centering
    \includegraphics[width=1.0\linewidth]{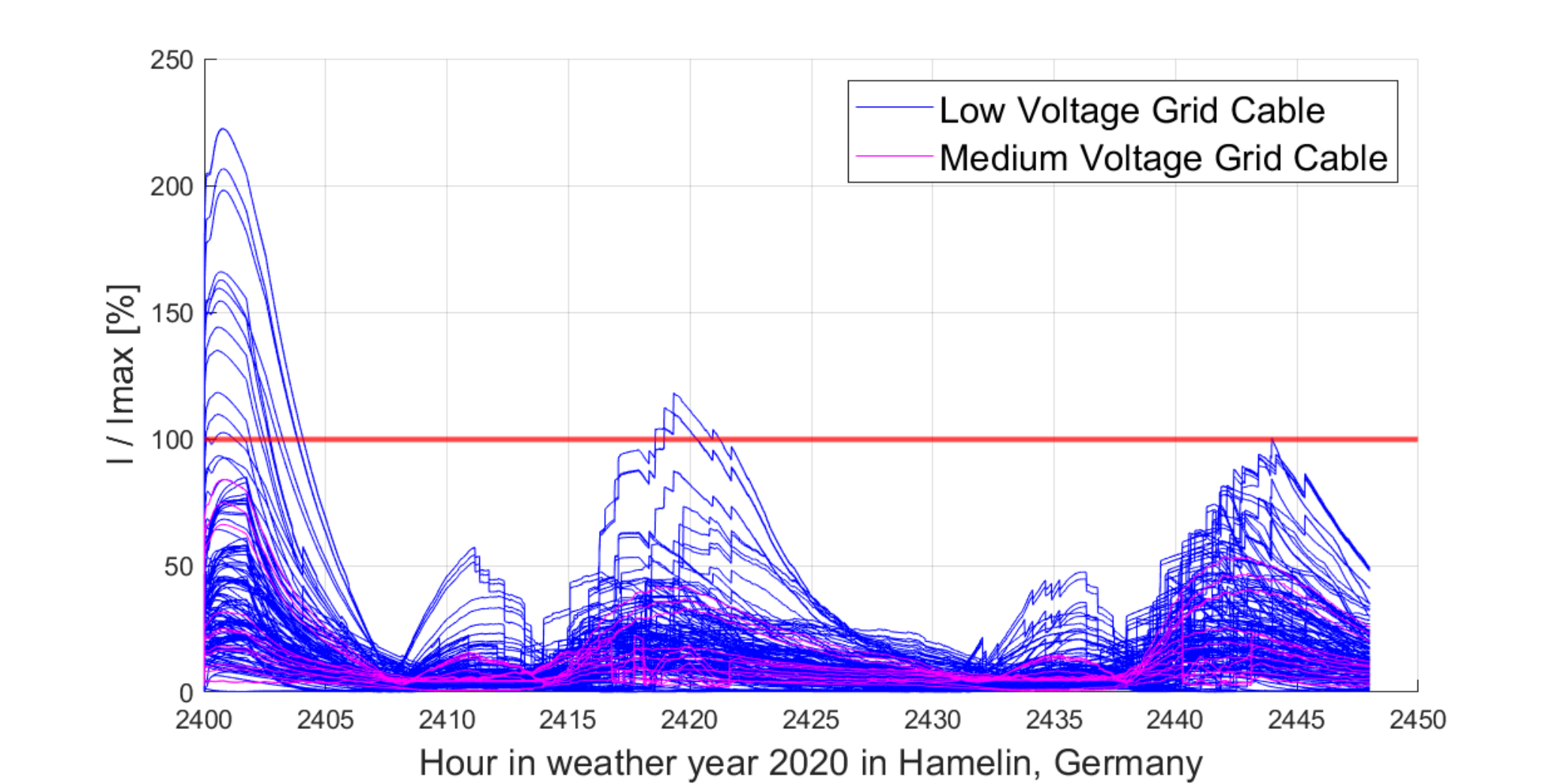}
    \caption{Line loading in percent based on the current-carrying capacity for low and medium voltage cables in the rural medium voltage ring scenario.}
    \label{fig:MV-Ring-Line}
\end{figure}

\begin{figure}[t]
    \centering
    \includegraphics[width=1.0\linewidth]{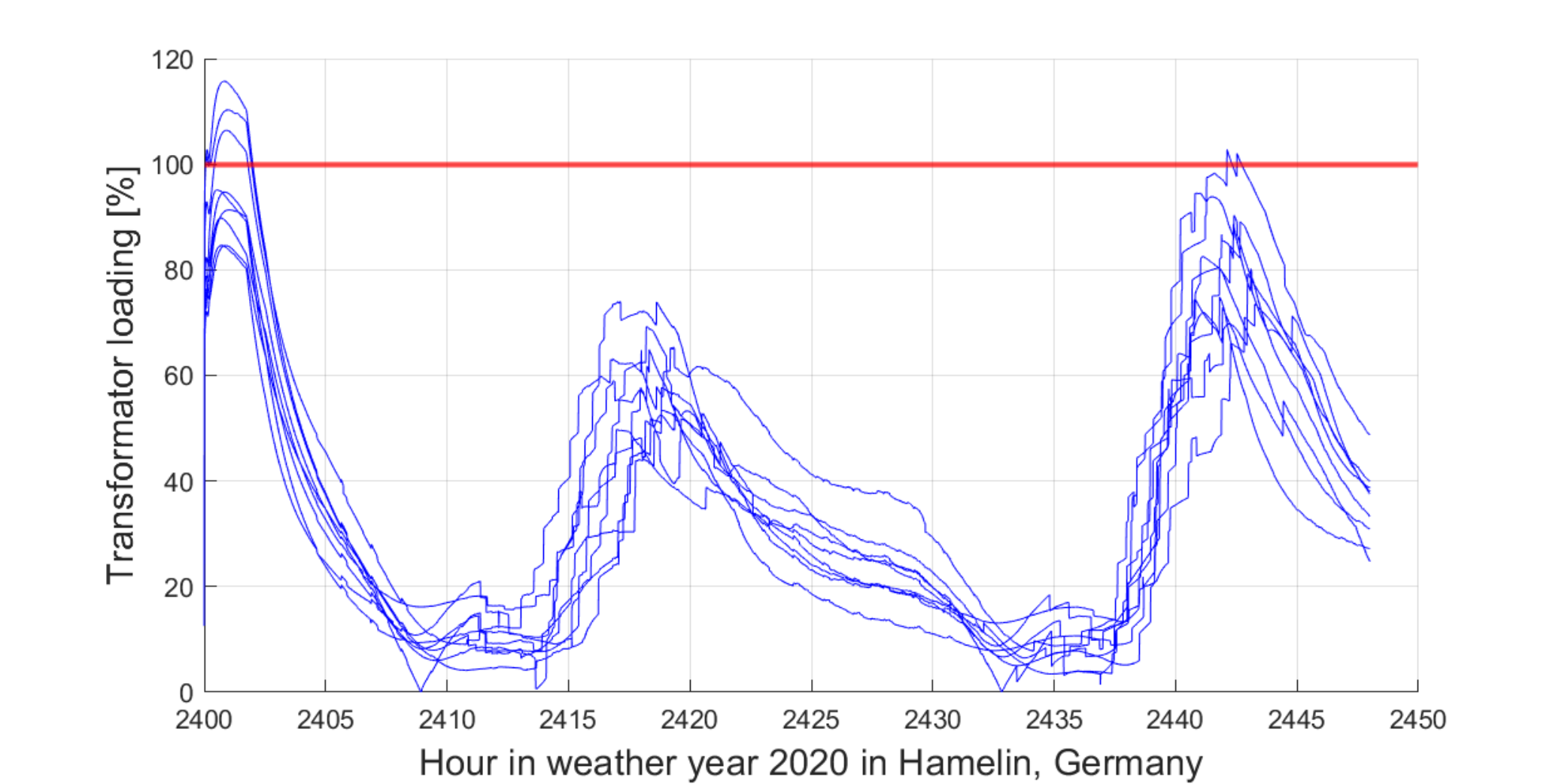}
    \caption{Transformer loading in percent for transformers between the low and medium voltage level in the rural medium voltage ring scenario.}
    \label{fig:MV-Ring-Transformer}
\end{figure}

When investigating the situation in this grid, by means of the bus voltages given in Fig. \ref{fig:MV-Ring-Voltage}, line loadings given in Fig. \ref{fig:MV-Ring-Line} and transformer loading given in Fig. \ref{fig:MV-Ring-Transformer}, within a 48-hour simulation based on the historical weather from April 2020, it can be seen that even in an scenario with air temperature of about 20°C and with high \ac{PV} generation, future medium and low voltage grids will likely face congestion situations on a daily bases. Especially in the evening hours, between 5 and 10pm, recurrently increasing power increases can be observed due to high share of \acp{BEV}.

\section{Conclusion and Outlook} \label{sec:conclusion} 

    In modern energy systems, different energy grids are coupled since renewable energy arises mostly as electricity, and the other energy grids (heat, gas and transportation) have to be supplied by the electricity sector, leading to the concept of integrated resp. multi-modal energy systems. The purpose of this paper is to describe a database and developed models to analyze the operation of an integrated power system at living quarter level. Since research is increasingly moving towards making data and models available, all developed algorithms and modeling tools are made open source.
    
    Public available data is used to create a qualitative scenario description of the physical energy system, which then is converted to a quantitative scenario description in different modules. The grid module parameterizes the energy grids electricity and heat, and the prosumer module parameterizes the relevant building energy components BES, EHP, BEV and PV, its thermodynamics and measuring devices. Finally, the model generation module transfers the database using the Modelica Language and TransiEnt-Library into a mathematical model. In a use case as an exemplary scenario the congestion situation in a low- and medium-voltage grid is analyzed.

    Due to the modular structure and high level of detail of the model description, users can build different scenarios and, for example, investigate the penetration of renewable energies, interactions of different controller structures or new operating strategies.

    In the future, aim is to further expand the described scenario generator and to specify and validate the created models within the framework of the ERIGrid~2 project.

\bibliographystyle{IEEEtran}
\bibliography{Bibliography}

\end{document}